\documentclass{iaus}
\usepackage{graphicx}

\title[Abundances of AGB stars and PNe] 
{The Chemical Composition of Red Giants, AGB Stars and Planetary Nebulae}

\author[Gustafsson \& Wahlin]   
{Bengt Gustafsson$^1$ and Rurik Wahlin$^1$}

\affiliation{$^1$Department of Astronomy and Space Physics, Uppsala University\break
Box 515, SE-751 20 Uppsala, Sweden\break email: bg@astro.uu.se\\[\affilskip]}

\pubyear{2006}
\volume{234}  
\pagerange{1--10}
\jname{Planetary Nebulae in our Galaxy and Beyond}
\editors{M. J. Barlow \& R. H. M\'endez, eds.}
\begin{document}

\maketitle

\begin{abstract}
The determinations of element abundances in red-giant stars and in particular in AGB stars are reviewed and
the resulting abundances are compared with those obtained for planetary nebulae in the Galaxy and
in nearby galaxies. The problems, possibilities and implications of such comparisons when estimating yields 
from low-mass and intermediate-mass stars are illustrated and commented on. 
\end{abstract}

\firstsection 
\section{Introduction}

Unlike the theory of stellar evolution, which is often characterized as one of the most successful theories
in astronomy, the theory of galaxy evolution is in an unsatisfactory state. The uncertainties as regards 
star-formation rates and IMF are reflected in free parameters, and also parameters specifying rates of infalling 
intergalactic gas and mixing of the ISM must be set. When it comes to yields from different types of stars 
other free parameters
must be specified, not the least for the contributions of intermediate- and low-mass stars. Basically, those
yields are the result of stellar-evolution calculations but these are, in spite of the general success of that theory,
not sufficient or reliable in this respect. Essential uncertainties
are stellar mass-loss rates as well as parameters related to the intricacies of mixing processes close to the regions of
nuclear processing in the stars. 
In this situation it would be very important to stabilize the ``theory'' 
by some solid observed yields, for SNe, Wolf-Rayet
stars, PNe and red giants (\cite{Gus04}). Such observations are, however, also uncertain and dependent on
unsatisfactory model atmospheres with additional free parameters. 

Attempts to systematically improve this situation, 
by increasing the accuracy of the model atmospheres and resulting abundance determinations 
for such objects to infer yields more directly, must be considered to have high priority, not only for learning more 
about effects of nucleosynthesis in the particular objects but for understanding galaxy evolution as a whole.

In the present review we shall focus on some aspects of such an ambitious programme: the determination of 
abundances in red giants, including AGB stars, and PNe. Due to space limitations, we shall concentrate 
on stars with initial masses below approximately $4$ M$_{\odot}$, i.e. stars that are not supposed
to undergo the Second dredge-up following core He burning, neither Hot Bottom Burning. This will also
exclude comparisons with PNe of Type I, rich in He and N, 
as well as with PNe with Wolf-Rayet central stars. As regards the recent development in chemical analyses of 
PNe, we refer to several contributions in the present volume, in particular those of Costa, Dinerstein, Liu, Peimbert, 
Shaw and Sterling.    


The chemical analysis of red-giant stars presents a number of difficult problems if one wants to reduce systematic
errors to a level where also moderate but important effects, e.g. due to internal mixing or dredge-ups, of a factor of two
or less in abundance may be traced. Recent observations from 
major telescopes and spectrometers are of such a high quality in terms of S/N that errors in the theoretical
analysis, not in the very observations, dominate the error budget for the abundance determinations,
at least for stars brighter than V $\sim 16$. In any analysis, first {\it stellar fundamental
parameters} such as T$_{\rm eff}$, acceleration of gravity at the stellar surface (or mass and radius of the star) must be
estimated. For the effective-temperature estimate, interstellar or circumstellar reddening or thermal dust
emission in the IR are major systematic sources of error, as well as the effects of uncertainties in the heavy molecular absorption
in the stellar photospheres. These may typically lead to errors of $5 \%$ in T$_{\rm eff}$ which is significant since
the abundance estimates may be very temperature sensitive, not least when derived from the strength of molecular lines.
The surface gravity estimate is similarly important -- it may often not be obtained accurately from the spectrum but
has to be determined from the apparent magnitude and an estimated distance, or just from an adopted absolute magnitude, 
and an assumed mass. For
galactic AGB stars, where the distances may still be quite uncertain as compared with stars in nearby galaxies, the error in
the surface gravity may lead to errors in absolute abundances of more than a factor of two, essentially reflecting the
pressure sensitivity of molecular equilibria. 
Still more problematic are the {\it errors in the model atmospheres}. The 
consequences of these problems for chemical analysis have not yet been fully explored. For the Sun, \cite{Asp05} have found
that proper consideration of the inhomogeneities and the cooling of the upper photosphere due to gas expansion
leads to revisions of CNO
abundances by about -0.2 dex. The only quantitative study of the
effects of adopting such more realistic 3D hydrodynamical models at abundance determinations for red giants is the work by  
\cite{Col06} on 
the extremely metal-poor giant HE0107-5340 ([Fe/H] = -5).  Collet et al. find that the CNO element abundances, based on diatomic
molecules, are reduced by a factor of 10 to 100 when convection is treated more realistically. Effects on abundances based on atomic lines
are significantly smaller, typically a factor of two. Although the corresponding effects are probably smaller for metal-rich stars
they may be significant. Studies on convection effects on spectra of red 
supergiants, using the ``star in a box'' models of 
(Freytag, in preparation) are presently being pursued. The effects of 
pulsations on spectra of Mira stars have been explored by
\cite{Gau04}. As regards the model effects of departures from LTE, as well as the uncertainties due
to missing or incomplete opacity data and dust formation in the outer stellar layers, they may also be quite considerable (for references,
see \cite{GusHof04}).
Many AGB stars have circumstellar molecular envelopes, and one could wonder whether their mm and sub-mm lines could
offer possibilities to obtain more accurate abundances. A short answer to 
that question is ``No''. Abundance estimates from these 
regions are judged to be accurate to, at best, a factor of 5 or so (\cite{Nym93}, \cite{Woo03}, \cite{Olo04}), and are sometimes
in error by one order of magnitude. Isotopic ratios may, however, be accurate to 25\% or even better for some sources, like the
famous dust-enshrouded carbon star IRC+10216.

\section{The first dredge-up as traced in observed abundances}

When empirically estimating yields of heavy
elements from low-mass and intermediate-mass stars, the dredge-ups and mass
loss before the AGB stages are important. We shall here
briefly summarize the observed evidence for dredge-up episodes before the 
AGB. 

The first episode, occurring already in the sub-giant 
phase, is predicted to bring CN processed material with depleted $^{12}$C and 
correspondingly enhanced $^{14}$N, as well as $^{13}$C, to the stellar surface.
The reduction of carbon and enrichment of nitrogen was qualitatively 
demonstrated for Pop I giants 
by \cite{LaRi77} and \cite{Kja82} but it was found by the latter
authors that the nitrogen enrichment was not large enough to match
carbon depletion. Recently this mismatch seems to have been unexpectedly 
resolved by the finding of Asplund et al. (2005) that the solar CNO 
abundances are
significantly lower than earlier believed; this reduces the carbon depletion,
and increases the nitrogen enrichment in the red giants, assuming that
they initially had solar-like abundances. The $^{13}$C enrichment, expected to
lead to $^{12}$C$/^{13}$C ratios of about 20, was early verified by \cite{Dea75} 
although some stars were also found with significantly lower ratios,
approaching the equilibrium CN-burning value of about 4. Empirically, one also
finds stars on the upper giant branch, starting around the bump in the 
luminosity function, which have depletions in carbon abundances and enhanced
nitrogen abundances significantly greater than those predicted (for a recent
discussion, see Gratton et al. 2000). \cite{SwMe79} earlier suggested that 
these 
phenomena might be due to meridional circulation driven by a rapidly spinning
stellar core; this may well be the case but, as was recently shown by 
\cite{Pal06}, the observed effects seem larger than those expected from
evolutionary models with angular-momentum transport and shear instibilities
taken into account. Perhaps more detailed models of convection-rotation 
interaction can produce larger effects. 

%


\section{The third dredge-up: galactic AGB stars vs PNe}

The third dredge-up is generally related to the thermal
pulses on the AGB where He is burnt in flashes in a thin shell. 
The following periodic rise of the temperature
gradient above the shell creates a convection zone bringing He-burning products
to higher layers from where the outer convection zone may later transport 
these products to the stellar surface. 

Although this general scenario is not strongly disputed, it is as yet not fully understood
quantitatively. When \cite{Ib81} formulated ``The Carbon Star Mystery'' 
this was reflecting the fact that the dredge-up of carbon only seemed to occur for
core masses of about 0.6 M$_{\odot}$ or more, while the observed luminosity 
function of carbon stars in the LMC indicated that lower-mass stars may
become carbon stars. This problem now seems 
solved for the LMC, while it still may be hard to obtain the observed low
lumnosities of many carbon stars in the SMC (\cite{Stan05}).

As regards the other characteristic consequences of the third dredge up, the enrichment
of the surface layers by s-elements, the situation is theoretically still more
unclear. Beautiful empirical studies, observing low Rb/Sr ratios and 
Mg and Zr isotopic ratios (\cite{Lam95}, \cite{Smi00}, \cite{Abi01}), have demonstrated that the
s enrichment is mainly the result of the relatively low neutron densities 
and the $^{13}$C($\alpha$,n)$^{16}$O source. However, the standard models
do not produce enough $^{13}$C in a hot enough environment. Some extra 
mixing leading to ingestion of protons into the intershell region would lead
to a $^{13}$C pocket (\cite{Gal98}) but also overlaying $^{14}$N which, due to
the high cross section of this nucleus for (n,p) reactions, could destroy the
neutrons. The blending of rotational mixing and convective overshooting 
in the bottom of the envelope convection zone with diffuse mixing is not
self-consistently handled yet, and the resulting s-element yields are 
dependent on free parameters, as well as on the angular momentum of the star
(cf. \cite{Her03}, \cite{GoSi04}, as well as the contribution by Busso to the 
present volume).

In this situation, where the resulting yields are dependent
on uncertain, or possibly not even physically fully adequate, parameters
in the stellar models, it is particularly important to obtain abundances
empirically. For the CNO elements and corresponding isotopes
the most complete and accurate studies are
still those by Lambert, Smith and collaborators (e.g., \cite{SmLa85}, \cite{SmLa86}, 
\cite{Lam86}, \cite{SmLa90}), partially built on analyses of high-resolution 
FTS spectra in the IR. These authors also compared
the CNO abundances of the stars with corresponding data for PNe (see 
in particular \cite{SmLa90}). In addition to a generally good agreement in
the abundance patterns, there are some interesting deviations from what might
be expected, in particular as regards the results for the 
N-type carbon stars (\cite{Lam86}): (1) The nitrogen abundances of these stars
are generally lower than those of the M and S stars; in fact N/O is
approximately solar, which would not be expected since the stars have gone
through the first dredge-up. This N/O distribution, shifted towards low N
abundances, is not reproduced by the carbon-rich PNe, even though there are
some nebulae that have lower N/O ratios than the M, MS and S stars tend
to show. A systematic uncertainty in the N abundances of Lambert et al. 
(1996) is due to
the dissociation energy of the CN molecule (providing the most important N-abundance
criteria) but recent new estimates of this quantity do not suggest an
important revision downwards which would be needed to reach
consistency with the PNe (cf \cite{Red03}). A more probable explanation may be 
errors in the model atmospheres. (2) The distribution of 
C/O ratios of the carbon stars shows a rather narrow peak at ratios between 1.0
and 1.4, while carbon-rich planetary nebulae show a much more extended 
distribution towards high C/O values, even beyond 2. This might reflect 
selection effects in the sample of Lambert et al.
(1996; the study is limited to
the apparently brightest objects in the K band, and dust enshrouded and presumably
quite carbon-rich objects are thus missed). We note that \cite{OhTs98b} have 
made an independent analysis of 3 of the stars analysed by Lambert et al.
(1996) and
obtain significantly higher C/O ratios, which seems to reflect differences
in model atmospheres and temperature scales.

Lambert et al. (1996) also provided $^{12}\mathrm{C}/^{13}\mathrm{C}$ 
ratios, based on
several different sets of photospheric molecular IR lines which were found to 
give mutually consistent results. The results were found to be consistent
with an increased C abundance due to a straight enrichment of $^{12}$C, with no
$^{13}$C added in the third dredge-up, excluding the small minority of 
$^{13}$C-rich carbon stars (``J-type stars''). Systematically smaller 
isotopic ratios were 
obtained for the same stars by \cite{OhTs96}, based on lines in more blended near-IR spectral
regions. The methods used by the latter authors were, however, criticised by 
\cite{deGu98}; see also \cite{OhTs98a} and \cite{deGu99}. 

\cite{ScOl00} and Woods et al. (2003) have derived 
$^{12}\mathrm{C}/^{13}\mathrm{C}$ ratios
for envelopes surrounding galactic carbon stars from mm CO lines. For a number of stars in 
common with Lambert et al. (1986) the agreement is good. As is seen in 
Fig. 1a, however, 
the distribution is different with a larger fraction of stars with lower 
isotopic ratios. If the stars with high and low mass-loss (\.M) are 
plotted separately there is also an indication of a steeper slope in the 
$10\lesssim\mathrm{^{12}C}/^{13}\mathrm{C}\lesssim40$ range for the high \.M 
sample. The result seems to suggest that the \cite{ScOl00} sample is
more biased towards low ratios, or possibly that there is a selective 
depletion of $^{12}$C relative to $^{13}$C in the envelopes. 

$^{12}\mathrm{C}/^{13}\mathrm{C}$ ratios
have recently been derived for PNe by several authors, primarily from mm lines 
(\cite{2005A&A...429..977W}, \cite{ 2002ApJ...572..326B}, \cite{ 2003A&A...397..659J}, 
\cite{ 1997A&A...324.1123B}, \cite{ 2000A&A...355...69P}, \cite{ 1988A&A...198L...1L}), 
but also from the hyperfine structure C~{\sc iii}] line at 1908~{\AA} 
(\cite{1997MNRAS.284..348C}).
The cumulative distribution is shown in Fig. 1a for carbon-rich PNe, as 
analysed by
\cite{ 1989Ap&SS.154...21A},  \cite{ 1989ApJ...345..306L}, \cite{1994Ap&SS.219..231M}, 
\cite{ 2003ApJ...585..475H}, \cite{ 2004A&A...423..199C}, \cite{ 2005MNRAS.362.1199C} 
and references therein, when abundances are not given in the same paper as the 
$^{12}\mathrm{C}/^{13}\mathrm{C}$. 
This distribution does, however, not depart significantly from the distribution
for all PNe with data available. 
It is seen that the distribution for PNe
again contains a greater fraction of stars with relatively low isotopic
ratios ($10\lesssim\mathrm{^{12}C}/^{13}\mathrm{C}\lesssim40$) than the distribution of photospheric 
ratios. This might be due to selection 
effects or selective depletion -- another possibility is systematic errors in
the analysis of the mm line data. We note, however, that the values of \cite{OhTs96} 
fit the PNe distribution rather well. 

\begin{figure}
\includegraphics[angle=0, width=0.5\textwidth]{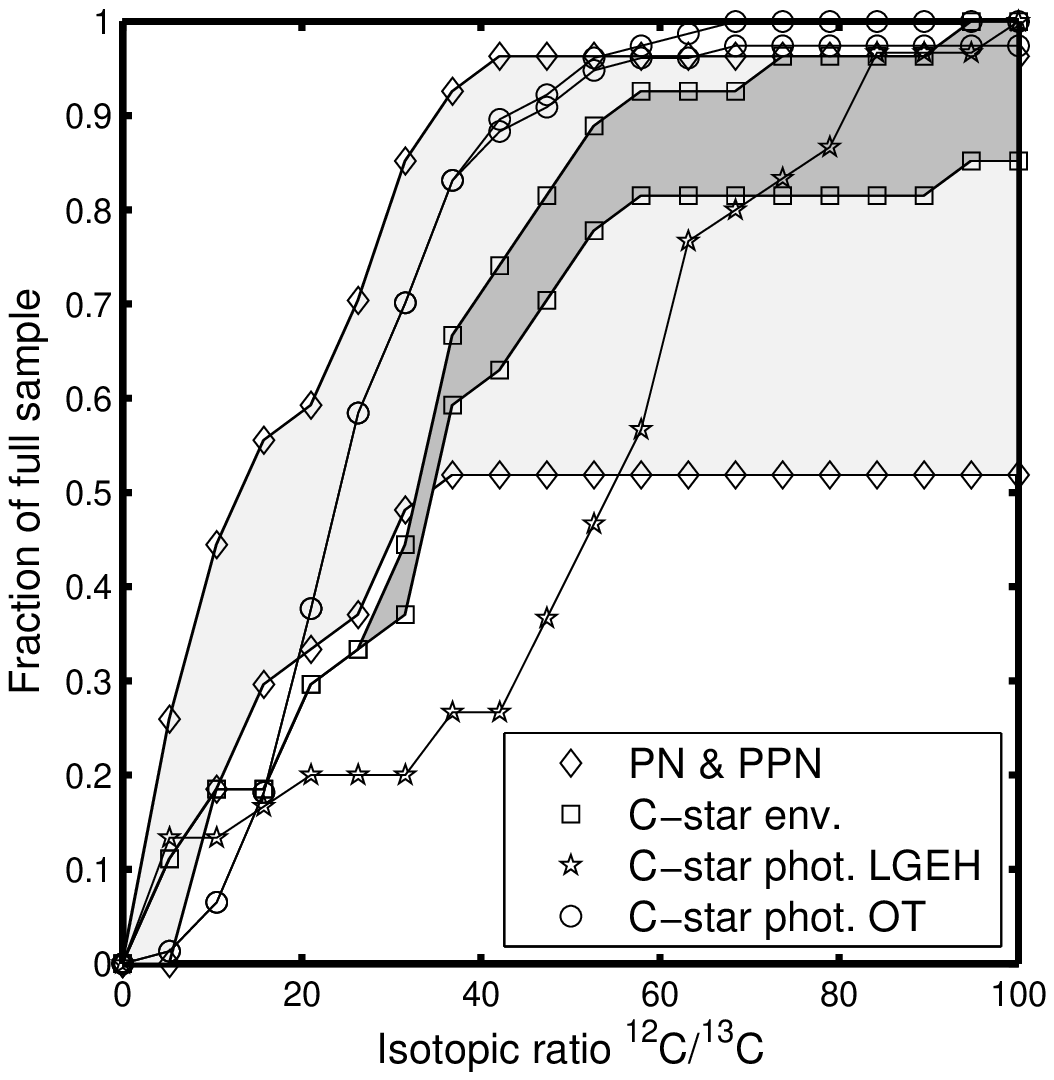}
\includegraphics[angle=0, width=0.5\textwidth, 
height=0.5\textwidth]{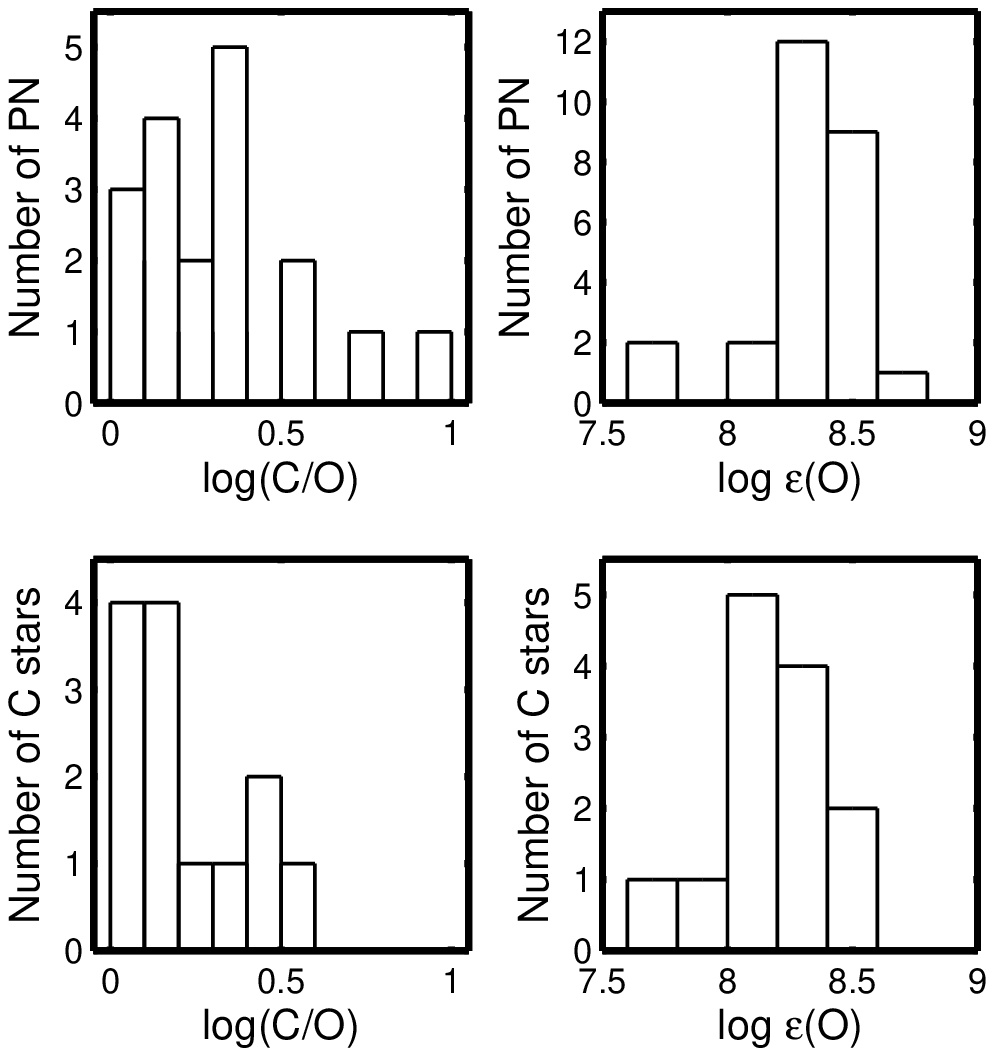}\\
\caption{(a; left): The cumulative distribution of 
$^{12}\mathrm{C}/^{13}\mathrm{C}$ is plotted for different samples. The
samples containing lower limit observations give a range of possible
values. In the PNe and PPNe sample only objects with C/O $>1$ are
included. LGEH represents Lambert et al. (1986), OT \cite{OhTs96}.  (b:
right): Histograms over C/O ratio and oxygen abundance for carbon rich PN
(Stanghellini et al. 2005) and carbon stars (\cite{Wah06}) in the LMC. The
oxygen abundances are in fair agreement with each other and the C/O ratio
is on average higher in the PN as expected.}
\end{figure}

There are several possible explanations for the discrepancy between PNe and typical carbon stars. 
There could be systematic errors in the observations or analysis giving
too low ratios in PNe and high \.M carbon shells or too high ratios in the
low \.M carbon stars.  Alternatively, the abundances might change when
normal carbon stars evolve to high \.M carbon stars to PNe. This could
either be a nuclear evolution producing $^{13}\mathrm{C}$ or possibly
destroying $^{12}\mathrm{C}$ or an error in the common assumption that
processes dissociating or regenerating $^{12}\mathrm{CO}$ or
$^{13}\mathrm{CO}$ in the circumstellar medium balance equally for the two
molecules (c.f. \cite{1988A&A...198L...1L}).  Another explanation could be
that carbon stars such as those observed by Lambert et al. (1986) would 
often not
be progenitors to typical C-rich PNe but instead end without a final
intensive wind, and if they produce PNe at all maybe mainly produce
spherical ones (cf. \cite{Sok02}). This would then also explain the
different morphologies of spherically symmetric shells observed around
some bright carbon stars (cf. \cite{Olo04} and references therein) as
compared with the normal non-spherical PNe. It should be noted that the
distribution relative to galactic longitude of the different samples is
not different enough to allow the conclusion that the discrepancy could
result from the radial gradient of $^{12}\mathrm{C}/^{13}\mathrm{C}$ in
the Galaxy (\cite{Mil05}).

Turning to the s-element abundances in PNe
\cite{Din01} has identified fine-structure transitions of Kr~{\sc iii} and
Se~{\sc iv} in the near IR spectra of planetary nebulae and \cite{Ste02} 
traced Ge~{\sc iii} far-UV lines in several PNe, and possibly even Ga~{\sc 
iii} (\cite{StDi03}). 
These elements are supposed to be produced early in the s-process. 
Enrichments of them with the same
range as for s-elements in AGB stars are found. Sterling and Dinerstein
suggest a variation of the abundances with C/O for Se and possibly
for Kr (see present proceedings, their Figure 1). When plotting the 
corresponding diagrams of the s-element abundances for AGB stars,
using the data of Smith \& Lambert (1985), Smith \& Lambert (1986) and 
Abia et al. (2002), we find
a slope for C/O $< 1.0$, much steeper than the line suggested by Sterling and 
Dinerstein (but in fact consistent with their data), but no clear increase in 
[s/Fe] for C/O $> 1.0$. Further work along these lines seems worthwhile.

\section{Local group AGB stars and PNe}
Important further constraints on the third dredge up, as well as on the
yields from intermediate- and low-mass stars in the early evolution of
galaxies, could be obtained from abundance analyses of AGB stars in 
Local Group galaxies. Here, the effects of differences in initial
abundances may be explored. One example of what may be done is the recent study by
\cite{deL06} of one carbon star in the SMC and two in the Sagittarius 
dwarf
spheroidal, representing stellar populations of different metallicity. 
The authors obtain abundances of various s elements from
high-resolution VLT/UVES spectra, and find that 
the ratio of high-mass s/low-mass s abundances 
(e.g. Ba, La, Ce and Pr, relative to Sr, Y and Zr) seems to vary with
metallicity in fair agreement with the variation predicted by Gallino et 
al. (1998)
with a constant value assumed for the mass of the $^{13}$C pocket.

In a study of about 50 carbon stars in several local-group dwarf galaxies 
(\cite{Wah06}) we observed spectra in the 1.6 and 2.2 $\mu$m bands
using the VLT/ISAAC
and the Phoenix IR spectrometer on Gemini South. 
We found the oxygen abundance of the program stars to
be enhanced on average, compared to the overall metallicity of the host galaxy, by
[O/Fe]$\sim0.25$. This is in fair agreement with the overabundance of
oxygen as measured in the ISM of such systems (\cite{Mat98}, \cite{Wes90}, \cite{Duf84}). 
The C/O ratio (by number) increases with decreasing oxygen
abundance and with increasing luminosity as
\begin{equation}
{\rm log(C/O) = -0.6 \cdot log \epsilon(O) - 0.3\cdot M_{bol} + 3.5,} 
\end{equation}

\noindent if the few $^{13}$C rich stars are excluded.
The $\epsilon$(O) dependence is roughly consistent with the yield 
estimates of \cite{Gav05}. 


The C/O ratios of the carbon stars in the LMC
are on average higher than the ratios in the galactic carbon stars studied
by \cite{Lam86}, but they are lower than the average C/O of LMC PNe 
as explored by \cite{Sta05} (see Fig. 1b).
Our preliminary estimates of C/O ratios of carbon stars in the
Sagittarius dSph are also lower than the C/O ratios of its PNe 
(\cite{Zij06}) and this also seems to be the case for the Fornax
galaxy.
The $^{12}\mathrm{C}/^{13}\mathrm{C}$ ratios were found to
be high in most carbon stars in the dwarf galaxies, as in their galactic correspondents.
The three carbon stars, in the Sagittarius dSph and SMC, studied by 
\cite{deL06} show $^{12}\mathrm{C}/^{13}\mathrm{C}>20$,
consistent with this picture.

Most stars with relatively high abundance of $^{13}\mathrm{C}$ were
found in two distinct areas in the HR diagram. Their M$_{\rm bol}$ and
$\mathrm{T}_{\rm eff}$ are consistent with (1) massive evolved
stars that could have been affected by hot-bottom burning and
(2) stars seemingly below the thermally pulsing AGB stars that 
underwent extrinsic carbon enrichment at low luminosity or were possibly
affected by a dredge-up at the He core flash.

Only three of our stars have reliable estimates of the nitrogen abundance,
although the uncertainty is probably about 0.5 dex. Two of these have
overabundances of nitrogen by [N/Fe]$\sim0.5$. One of them is also enhanced 
in oxygen which may indicate an Fe abundance higher than assumed on the basis of the
overall metallicity of the galaxy. 
The third star has an overabundance of [N/Fe] $\sim1$. This is probably high
enough to be significant and the  $^{12}\mathrm{C}/^{13}\mathrm{C}\sim4$
indicates that the gas in the star has been CN-processed.

\section{Conclusions and prospects}
It should be clear from the discussion above that contemporary results of
abundance determinations for AGB stars and for PNe permit interesting 
comparisons,
though still not many firm conclusions. It seems that oxygen abundances 
of carbon stars and carbon-rich PNe have not been very
much affected by intrinsic nucleosynthesis but sooner measure the initial abundances.
Also, the shift of the carbon-abundance distribution for carbon-rich PNe relative
to the carbon stars indicates that the carbon enrichment proceeds and in many cases 
reaches its highest value in the superwind, ending the AGB phase and initiating the 
PN phase. This is supported also by the observed increase of the C/O ratio as a function of
carbon-star luminosity. The observed tendency for C/O in carbon stars 
to increase with decreasing initial metallicity is also expected.  There are still unsolved 
problems as regards the unexpectedly low N abundances obtained for
the galactic carbon stars, as well as their high $^{12}\mathrm{C}/^{13}\mathrm{C}$ ratios 
when compared with PNe; these discrepancies might, however, be due to systematic errors in the
analyses. The analysis of AGB star abundances have given very important clues for the understanding
of the s process; there is also a general overall consistency between these abundances
in PNe and AGB stars, but more data are needed. 

In addition to methodological improvements in abundance analysis for stars and PNe, efforts to
measure abundances of new elements are important; one obvious example is sulphur in carbon stars, e.g. using CS and CO lines
to study the S/O ratio. Also, the possibilities to systematically exploit new wavelength regions, not the least
the near and intermediate IR both for stars and PNe, are presently developing and still very promising.

\begin{acknowledgments}
Kjell Eriksson, Susanne H\"ofner and Nils Ryde are thanked for valuable comments on the manuscript.
Albert Zijlstra is thanked for providing data on PNe abundances in the Sagittarius
and Fornax systems before publication. 

\end{acknowledgments}

\end{document}